\documentclass[twocolumn,showpacs,prb]{revtex4}

\usepackage{graphicx}
\usepackage{dcolumn}
\usepackage{bm}

\begin{document}

\title{Microscopic theory of quantum dot interactions with quantum light: local field
effect}


\author{G.Ya. Slepyan}%
\author{A. Magyarov }
 \email{andrei.magyarov@gmail.com}
\author{S.A. Maksimenko }

\affiliation{%
Institute for Nuclear Problems, Belarus State University,
Bobruiskaya 11, 220050 Minsk, Belarus
}%
\author{A. Hoffmann}
\affiliation{\\ Institut f\"{u}r Festk\"{o}rperphusik, Technische
Universit\"{a}t Berlin, Hardenbergstr. 36, 10623 Berlin, Germany}


\begin{abstract}
A theory of  both linear and nonlinear electromagnetic response of a
single QD exposed to quantum light, accounting the depolarization
induced local--field has been developed. Based on the microscopic
Hamiltonian accounting for the electron--hole exchange interaction,
an effective two--body Hamiltonian has been derived and expressed in
terms of the incident electric field, with a separate term
describing the QD depolarization. The quantum equations of motion
have been formulated and solved with the Hamiltonian for various
types of the QD excitation, such as Fock qubit, coherent fields,
vacuum state of electromagnetic field and light with arbitrary
photonic state distribution.  For a QD exposed to
coherent light, we predict the appearance of two oscillatory regimes
in the Rabi effect separated by the bifurcation. In the first
regime, the standard collapse--revivals phenomenon  do not reveal
itself and the QD population inversion is found to be negative,
while  in the second one, the collapse--revivals picture is found to
be strongly distorted as compared with that predicted by the
standard Jaynes-Cummings model.
For the case of QD interaction with  arbitrary quantum light state
in the linear regime, it has been shown that the local field induce
a fine structure of the absorbtion spectrum. Instead of a single
line  with frequency corresponding to which the exciton transition
frequency, a duplet is appeared with one component shifted by the
amount of the local field coupling parameter. It has been
demonstrated the strong light--mater coupling regime arises in the
weak-field limit. A physical interpretation of the predicted effects
has been proposed.

\end{abstract}

\maketitle

\section{Introduction}
 The strong coupling between condensed matter and quantum light
is a core issue of present day quantum optics. Realized by exposing
the matter with intense quantum field, it can manifest itself  in
different quantum systems such as single atoms and ultracold atomic
beams \cite{Scully}, semiconductor heterostructures
\cite{bimberg_b99,Michler_book}, Bose--Einstein condensates
\cite{kasevich}, \textsl{etc.} Albeit these systems are of different
physical nature, their interaction with quantum light is governed by
common rules. In the strong coupling regime, these systems enable
the generation of different states of quantum light --- single
photons \cite{lounis_05}, Fock states \cite{bratke}, Fock qubits,
quantum states with arbitrary photon number distribution
\cite{Law_96}. That constitutes a basis for the quantum information
processing \cite{Kilin_99,zadkov} and quantum metrology
\cite{wheeler}. In practice, the strong light--matter coupling
regime  can be realized in two ways: by combining the matter with a
high-Q microcavity or by exposing the matter to a ultrashort intense
pump pulse.

To describe the strong coupling  between  an arbitrary two--level
system and quantum light, the Jaynes--Cummings (JC) model is
conventionally used \cite{Jaynes_Cummings}. One of the most
fundamental phenomenon predicted within the JC model is the
oscillation of the population between levels with the Rabi frequency
(Rabi oscillations).  However, the standard JC model does not
account for a number of physical factors, which, under certain
conditions, may significantly influence the Rabi effect.
The time--domain modulation of the field--matter coupling constant
\cite{Law_96,yang_04} and interplay between classical driving
field and quantized cavity field \cite{Law_96} can serve as
examples. More advanced JC models involve additional interaction
mechanisms and effects, such as dipole--dipole (d--d) interaction
\cite{lewenstein_94, Zhang_94}, exction--phonon coupuling
\cite{forstner_03,dizhu_05}, and self--induced transparency
\cite{fleischhauer_05}. The d--d interaction between two quantum
oscillators leads to radiative coupling of them and, as a result,
to exchange by the excited state. That is,  Rabi oscillations
between these two oscillators occur; see Ref. \onlinecite{Dung_02}
for a theory and Ref. \onlinecite{Unold_05} for the
experimental observation in a double quantum dot (QD) system. As a
whole, the observation and intensive studying of excitonic Rabi
oscillations
\cite{Unold_05,sticvater,kamada,htoon,Zrenner_nature,sticvater_rep,mitsumori_05}
motivates the  extension of the JC--model to incorporate specific
interactions inherent to confined exciton in a host.

In the given paper we  present a microscopic theory of the
interaction of an isolated QD with quantum light for both weak and
strong coupling regimes. We incorporate the local field correction
into the JC model as an additional physical mechanism influencing
the Rabi effect in a QD exposed to quantum light. In particular, the
Rabi oscillations are shown to exist even in the limit of a weak
incident field.

In the weak coupling regime, the local--field effects in optical
properties of QDs have been theoretically investigated in Refs.
\onlinecite{Schmitt_87,Hanewinkel_97,Slepyan_99a,Maksim_00a,Ajiki_02,Goupalov_03,Slepyan_NATO_03,Maksimenko_ENN,Maksimenko_HN04}
for classical exposing light  and in Ref.
\onlinecite{maxim_pra_02} for quantum light. In the latter case it
has been shown that for a QD interacting with Fock qubits the
local fields induce a fine structure of the absorption (emission)
spectrum: instead of a single line with the frequency
corresponding to the exciton transition, a doublet appears with
one component shifted to the blue (red). The intensities of
components are completely determined by the quantum light
statistics. In the limiting cases of classical light and Fock
states the doublet is reduced to a singlet shifted in the former
case and unshifted in the latter one.

The role of local fields in the excitonic Rabi oscillations in an
isolated QD driven by classical excitation was investigated in Ref.
\onlinecite{magyar04}. Two different oscillatory regimes separated
by the bifurcation have been predicted to exist. The Rabi
oscillations were predicted to be non-isochronous and arising in the
weak excitation regime. Both peculiarities have been experimentally
observed  by Mitsumori \textit{et al.} in Ref.
\onlinecite{mitsumori_05} where the Rabi oscillations of excitons
localized to quantum islands in a single quantum well were
investigated.

There exist several different physical interpretations of local
field in QDs and, correspondingly, different ways of its theoretical
description.  The first model (scheme A in the terminology of Ref.
\onlinecite{Ajiki_02}) exploits the standard electrodynamical
picture: by virtue of external field screening by charges induced on
the QD surface (the quasistatic Coulomb electron--hole interactions)
a depolarization field is formed differentiating the local (acting)
field in the QD and external incident field. In this model the total
electromagnetic field  is not pure transverse. Alternatively, only
transverse component is attributed to the electromagnetic field,
while the longitudinal component is accounted for through the
exchange electron--hole interactions (scheme B accordingly Ref.
\onlinecite{Ajiki_02}). Both approaches are physically equivalent
and lead to identical results.

In the present paper we build the analysis on the general
microscopic quantum electrodynamical (QED) approach where the local
field correction originates from the exchange by virtual vacuum photons
between electrons and holes forming the exciton and thus
is a manifestation of the dipole--dipole (d--d) interaction between
electrons and holes (the dynamical Coulomb interaction) \cite{lewenstein_94,Zhang_94}.
The approach allows us to overcome a number of principal
difficulties related to the field quantization in QDs \cite{ref01}.
In the analysis, approximate solution of the many--body problem is built on the  Hartree--Fock--Bogoliubov
self--consistent field concept \cite{lewenstein_94}. The self-consistent
technique leads to a separate term in the effective Hamiltonian responsible for the interaction of
operators and average values of physical quantities. Due to this
term the quantum mechanical equations of motion become nonlinear and require numerical integration.

The paper is arranged as follows. In Sect. \ref{sec:theory} we
develop theoretical model describing the QD--quantum light
interaction. We formulate a model Hamiltonian with the separate
term  accounting for the local field correction and corresponding
equations of motion. In Sect. \ref{sec:free_mot} we analyze the
manifestation of local fields in the motion of the QD exciton  in
the absence of external field. In Sec. \ref{sec:weak_f_lim} we
investigate the QD interaction with arbitrary state of quantum
light in the weak driving field regime. Sect.
\ref{sec:rabioscillations} is devoted to the theoretical analysis
of local field influence on the Rabi oscillations  in the QD
exposed to coherent states of light and Fock qubits. A discussion
of the results obtained is presented in Sect. \ref{sec:discussion}
and concluding remarks are given in  Sect. \ref{sec:conclusion}.

\section{Quantum Dot -- quantum  light Interaction: theoretical  model}

\label{sec:theory}

\subsection{Interaction Hamiltonian}

In this section  we formulate  the interaction Hamiltonian for a QD exposed to quantized field accounting for the local--field correction. Later on we exploit
the Hamiltonian for the derivation of equations of motions describing dynamical properties of this system.

As aforementioned, the local field in QD differs from the incident
one  due to the d--d electrons--holes interaction. A general
formalism accounting for the d--d interactions in atomic
many--body systems exposed to photons has been developed in Refs.
\onlinecite{lewenstein_94,Zhang_94} as applied to nonlinear optics
of Bose--Einstein condensates \cite{lewenstein_94,Zhang_94}.  We
extend this formalism to the case of the QD exciton driven by
quantized light.

Consider an isolated QD exposed to quantized electromagnetic field.
The electron--hole pairs in QD are assumed to be strongly confined;
thus we neglect the static Coulomb interaction between electrons and
holes. We decompose the operator of the total electromagnetic field
into two components. The first one,
$\widehat{\bm{\mathcal{E}}}_\mathrm{v}$, represents a set of modes
that do not contain  real photons. The second component,
$\widehat{\bm{\mathcal{E}}}_\mathrm{0}$, represents the set of modes
emitted by the external source of light (real photons). Such a
decomposition as well as the subsequent separate consideration of
the field components is analogous to the Heisenberg--Langvein
approach in the quantum theory of damping, see Ref.
\onlinecite{Scully}. The total Hamiltonian of the system
''QD+electromagnetic field'' is then represented as
\begin{equation}
\widehat{{\cal H}}=\widehat{{\cal H}}_\mathrm{0}+\widehat{{\cal
H}}_\mathrm{ph}+\widehat{{\cal H}}_\mathrm{vac}+\widehat{{\cal
H}}_\mathrm{I0}+\widehat{{\cal H}}_\mathrm{Iv} \, ,
 \label{init_Ham}
\end{equation}
where $\widehat{{\cal H}}_\mathrm{0,ph,vac}$ are the Hamiltonians
of the QD free charge carriers, the incident photons and the
virtual vacuum photons, respectively. The terms $\widehat{{\cal
H}}_\mathrm{I0, Iv}$ describe the interaction of electron--hole
pair with incident quantum field
$\widehat{\bm{\mathcal{E}}}_\mathrm{0}$ and with vacuum field
$\widehat{\bm{\mathcal{E}}}_\mathrm{v}$, respectively. In the
dipole approximation these Hamiltonians are given by
\begin{equation} \label{HamIl_HamIv}
    \widehat{{\cal H}}_\mathrm{I0,Iv}^{}=-\frac{1}{2}\int_{V}(\widehat{\bm{\mathcal{P}}}\widehat{\bm{\mathcal{E}}}_\mathrm{0,v}+
    \widehat{\bm{\mathcal{E}}}_\mathrm{0,v}\widehat{\bm{\mathcal{P}}})\, d^3\mathbf{r}\,
\end{equation}
where $V$ is the QD volume and
$\widehat{\bm{\mathcal{P}}}(\bm{r},t)$ is the QD polarization
operator. The Hamiltonian $\widehat{{\cal H}}_{\mathrm{vac}}$ is
as follows
\begin{equation}
 \widehat{{\cal H}}_{\mathrm{vac}}=\sum_{k\lambda}\hbar\omega_{k}\hat{v}^\dag_{k\lambda}\hat{v}_{k\lambda}\,,
\label{Ham_Hv}
\end{equation}
where $\hat{v}^\dag_{k\lambda}$ and $\hat{v}_{k\lambda}$ are the
creation and annihilation operators of vacuum photons, $k$ is the
mode index, indexes $\lambda=1,2$ denote the field polarization. The
operator of vacuum electromagnetic field
$\widehat{\bm{\mathcal{E}}}_\mathrm{v}$ is determined as
\begin{equation}
\widehat{\bm{\mathcal{E}}}_\mathrm{v}=i\sum_{k\lambda}\sqrt{\frac{2\pi\hbar\omega_{k}}{\Omega}}{\,}
\textbf{e}_{k\lambda}(\hat{v}_{k\lambda}{e}^{i\textbf{kr}}-\hat{v}^\dag_{k\lambda}{e}^{-i\mathbf{kr}})\,,
\label{field_operator_Ev}
\end{equation}
where $\Omega$ is the normalization volume and
$\mathbf{e}_{k\lambda}$ is the polarization unit vector.

As a first step in the development of our theory we eliminate from
the consideration the exchange interactions and proceed to the
direct interactions. For that purpose, we exclude the vacuum photon
operators $\hat{v}_{k\lambda}$ and $\hat{v}^\dag_{k\lambda}$
expressing them (and corresponding Hamiltonians  $\widehat{{\cal
H}}_\mathrm{vac} $ and $\widehat{{\cal H}}_\mathrm{Iv}$) in terms
of the polarization operator $\widehat{\bm{\mathcal{P}}}$.
Recalling the Heisenberg equation
$i\hbar\partial{\hat{v}_{k\lambda}}/\partial{t}=-\left[\widehat{{\cal
H}},\hat{v}_{k\lambda}\right]$, the expression as follows can be
obtained,
\begin{equation}
\frac{\partial{}\hat{v}_{k\lambda}}{\partial{t}}=-i\omega_{k}\hat{v}_{k\lambda}+\widehat{F}_{k\lambda}(t)\,,
\label{dif_eq_operator_B}
\end{equation}
where
$\widehat{F}_{k\lambda}(t)=\sqrt{2\pi\omega_{k}/\hbar{\Omega}}\int\limits_{V}
\widehat{\bm{\mathcal{P}}}(\textbf{r}',t)\textbf{e}_{k\lambda}e^{i\textbf{kr}}d^3{\textbf{r}'}$.
Solution of Eq. (\ref{dif_eq_operator_B}) is given by
\begin{equation}
\hat{v}_{k\lambda}(t)=\hat{v}_{k\lambda}(-\infty)e^{-i\omega_kt}+\int\limits_{-\infty}^{t}\!\widehat{F}_{k\lambda}(\tau)e^{-i\omega_k(t-\tau)}d\tau\,
\label{eq_operator_B}
\end{equation}
with the first term describing the free evolution of the reservoir modes (quantum noise) and the second one responsible for the
exchange interactions. Further we neglect the first term in (\ref{eq_operator_B}) leaving the quantum noise beyond the consideration. Inserting
then this equation into (\ref{field_operator_Ev}) and the resulting expression -- into Hamiltonian (\ref{HamIl_HamIv}), after some algebra we arrive at
\begin{eqnarray}
\widehat{{\cal H}}_\mathrm
{Iv}=-\sum_{k}\frac{2\pi{i}\omega_{k}}{\Omega}\int\limits_{-\infty}^{t}\!\!\int\limits_{V}\!\!\int\limits_{V}\widehat{{\mathcal
    P}}_{\alpha}(\textbf{r},t)\widehat{{\mathcal P}}_{\beta}(\textbf{r}',\tau)~~~~~~~~\cr\rule{0in}{4ex}
   \sum_{\lambda}{\bf \rm e}^{(\alpha)}_{k\lambda}{\bf \rm e}^{(\beta)}_{k\lambda}e^{i
\rm{\bf k}({\bf r}-{\bf r}')}e^{i\omega_{k}(\tau-t)}d\tau{}d^3{\bf r}\, d^3{\bf r}' +\rm{ H.c.}\,, \label{HamIl}
\end{eqnarray}
where indexes $\alpha$, $\beta$ mark Cartesian projections of vectors.  The summation over repetitive indexes is
assumed. Using the relationship \cite{Scully}
\begin{eqnarray}
\sum_{\lambda}{\rm e}^{(\alpha)}_{k\lambda}{\rm
e}^{(\beta)}_{k\lambda}&=&
\delta_{\alpha\beta}-\frac{k_{\alpha}k_{\beta}}{k^2}\cr&=&\frac{1}{k^2}\left(\frac{\partial^2}{\partial{x_{\alpha}}\partial{x_{\beta}}}-\delta_{\alpha\beta}\frac{1}{c^2}\frac{\partial^2}{\partial{t^2}}
\right)\nonumber\,,
\end{eqnarray}
we proceed  to the limit $\Omega\rightarrow\infty$ in (\ref{HamIl}). That corresponds to the replacement
\begin{eqnarray}
\sum_{k}[\cdot]\rightarrow\frac{\Omega}{(2\pi)^3}\int{[\cdot]}d^3\textbf{k}\,. \nonumber
\end{eqnarray}
Then, utilizing the Markov property of the polarization operator,
$\widehat{{\mathcal P}}_{\alpha}(\textbf{r},t)\simeq
\widehat{{\mathcal P}}_{\alpha}(\textbf{r},0)$
[\onlinecite{Zhang_94}], the Hamiltonian (\ref{HamIl}) is reduced
to
\begin{eqnarray}
\label{Ham_HIl_final} \widehat{{\cal H}}_\mathrm{Iv}&=&
    -4\pi\int\limits_{0}^{\infty}\!\!\int\limits_{V}\!\!\int\limits_{V}
    \left(\frac{\partial^2}{\partial{x_{\alpha}}\partial{x_{\beta}}}
    -\delta_{\alpha\beta}\frac{1}{c^2}\frac{\partial^2}{\partial{{t'}^2}}\right)~~~~ \cr \rule{0in}{4ex}
    &\times & G^{(0)}(\textbf{r}-\textbf{r}',t')\widehat{{\mathcal P}}_{\alpha}(\textbf{r},t)
    \widehat{{\mathcal P}}_{\beta}(\textbf{r}',t)d{t'}d^3\textbf{r}d^3\textbf{r}' \,,
\end{eqnarray}
where
\begin{eqnarray}
 G^{(0)}({\bf
r},t)&=&\frac{i c^2}{2(2\pi)^3}\int\frac{e^{i{\bf
kr}}}{\omega_k}\left( e^{-i\omega_kt}-e^{i\omega_kt}\right)d^3{\bf
k}\cr \rule{0in}{5ex} &=&\frac{1}{4\pi|{\bf
r}|}\left[\delta\left(\frac{|{\bf r}|}{c}
-t\right)-\delta\left(\frac{|{\bf r}|}{c}
+t\right)\right]\,,\label{green_funct}
\end{eqnarray}
is the free-space Green function \cite{beresteckij} and $\delta(\dots)$ is the Dirac
delta--function.
Evaluation of $\widehat{{\cal H}}_\mathrm{vac}$ in (\ref{Ham_Hv})
is carried out analogously and gives $\widehat{{\cal
H}}_\mathrm{vac}=-\widehat{{\cal H}}_\mathrm{Iv}/2$.

As the next step, we adopt the quasi--static approximation, which
utilizes the property of QD to be electrically small. The approximation implies
the limit transition $c\rightarrow\infty$ and the neglect the
terms $\sim\partial^2/\partial{t'}^2$ in Hamiltonian
(\ref{Ham_HIl_final}). Then, the Hamiltonians $\widehat{{\cal
H}}_{\mathrm{vac}}$ and $\widehat{{\cal H}}_{\mathrm{Iv}}$ are
represented by the sum as follows
\begin{eqnarray}
\label{deltaH} \Delta{\widehat{{\cal H}}}&=&\widehat{{\cal
H}}_{\mathrm{vac}}+\widehat{{\cal H}}_{\mathrm{Iv}}\cr
\rule{0in}{5ex}&=&-\frac{1}{2}\int\limits_{V}\!\!\int\limits_{V}\!\!
    \widehat{\bm{\mathcal{P}}}(\mathbf{r})
        \underline{G}(\mathbf{r}-\mathbf{r}')
    \widehat{\bm{\mathcal{P}}}(\mathbf{r}')
    \,d^3\mathbf{r}\,d^3\mathbf{r}'
\,,
\end{eqnarray}
where
\begin{equation}
\label{Green_Tens} \underline{G}({\bf r}-\mathbf{r}')=
    \nabla_\mathbf{r}\otimes\nabla_\mathbf{r}
    \left(\frac{1}{|\mathbf{r}-\mathbf{r}'|}\right)
    \end{equation}
is the free space Green tensor; $\nabla_\mathbf{r}\otimes\nabla_\mathbf{r}$ is  the operator
dyadic acting on variables $\mathbf{r}$. In the quasi--static
approximation we neglect the line broadening due to the dephasing
and the spontaneous emission. The latter effect can be introduced in the model by
retaining terms $O(1/c)$ in the the quasi--static approximation.

In the preceding analysis we have suggested that the exchange  by
virtual photons of all modes occurs between  all allowed dipole
transitions. That is, on that stage the problem was stated as a
quantum--mechanical many--body problem. The analysis can be
significantly simplified if we restrict ourselves to the two--level
approximation assuming the exciton transition frequency to be
resonant with the  acting field carrier frequency and utilize the
self--consistent field model. The self--consistent field is
introduced by means of the Hartree--Fock--Bogoliubov approximation
\cite{hartree_rigourus}, which implies the linearization of
Hamiltonian (\ref{deltaH}) by the substitution
\begin{equation}
\widehat{ {\bm {\mathcal P}}  }({\bf r})\widehat{{\bm {\mathcal
P}}}({\bf r}')\rightarrow\widehat{{\bm{\mathcal  P}}}({\bf
r})\langle\widehat{{\bm {\mathcal P}}}({\bf r}')\rangle
+\langle\widehat{{\bm {\mathcal P}}}({\bf r})\rangle\widehat{{\bm
{\mathcal P}}}({\bf r}')\,. \label{substitution_linearization}
\end{equation}
The polarization operator of two-level system is
given by \cite{Cho_b03}
\begin{equation}
\widehat{{\bm {\mathcal  P}}}({\bf r})=|\zeta({\bf r})|^2({\bm
\mu}\hat{\sigma}_{+}+{\bm \mu}^{*}\hat{\sigma}_{-})\,,
\label{polarization_operator}
\end{equation}
where $\hat{\sigma}_{\pm}$ are the Pauli pseudospin operators and
$\zeta({\bf r})$ is the wavefunction  of the electron--hole pair. In
the strong confinement regime this function is assumed to be turns
out to be the same both in excited and ground states
\cite{Chow_b99,haug_b94}.

In the two--level approximation, the Hamiltonian of the carriers
motion is represented as
\begin{equation}
\widehat{{\cal H}}_{0}=\varepsilon_{\mathrm{e}}\hat{a}_{\mathrm
{e}}^{\dag}\hat{a}_{\mathrm{e}}+\varepsilon
_{\mathrm{g}}\hat{a}_{\mathrm{g}}^{\dag}\hat{a}_{\mathrm {g}}\,,
\label{Ham_free}
\end{equation}
where $\varepsilon_{\mathrm{g},\mathrm{e}}$ and
$\hat{a}_{\mathrm{g,e}}^{\dag}$/$\hat{a}_\mathrm{{g,e}}$ are the
energy eigenvalues and creation/annihilation operators of the
exciton; indices $\mathrm{ e} $ and $\mathrm{ g} $ correspond to
the excitonic excited and ground states, respectively. The acting
field operator is expressed by the relation
(\ref{field_operator_Ev}) after the substitutions
$\hat{v}_{k\lambda}\rightarrow{}\hat{c}_{q}(t)$ and
$\hat{v}^\dag_{k\lambda}\rightarrow{}\hat{c}^\dag_{q}(t)$;
$\hat{c}^\dag_{q}(t)/\hat{c}_{q}$ are the creation/annihilation
operators of the incident (real) photons (the polarization index
$\lambda$ is included in the mode number $q$). Formally, the
relation  (\ref{eq_operator_B}) is fulfilled for operators
$\hat{c}_{q}(t)$ and $\hat{c}^\dag_{q}(t)$ too, and the first term
describes the evolution of real photons. However, since the
exchange interaction is included into the vacuum field component,
in the case of real photons the second term in  relation
(\ref{eq_operator_B}) disappears. Then, the Hamiltonian
$\widehat{{\cal H}}_{\rm{ph}}$ is given by the relation
(\ref{Ham_Hv}) after the substitution
$\hat{v}_{k\lambda}\rightarrow{}\hat{c}_{q}(-\infty)$ and
$\hat{v}^\dag_{k\lambda}\rightarrow\hat{c}^\dag_{q}(-\infty)$. For
shortness, we denote  $\hat{c}_{q}(-\infty)=\hat{c}_{q}$ and
$\hat{c}^\dag_{q}(-\infty)=\hat{c}^\dag_{q}$ .

Note that nonresonant transitions can be approximately accounted
through a real--valued frequency--independent background
dielectric function $\epsilon_h$.  Assuming $\epsilon_h$ to be
equal to dielectric function of surrounding medium, we put further
$\epsilon_h=1$ without loss of generality. Substitutions in final
expressions $c\to c/\sqrt{\epsilon_h}$ and ${\bm \mu}\to {\bm
\mu}/\sqrt{\epsilon_h}$ for the speed of light and the
electron-hole pair dipole moment, respectively, will restore the
case $\epsilon_h\neq 1$.

As a next step we in\-tro\-duce the ro\-tat\-ing wave approximation
\cite{Scully}, i.e., we  neglect  in (\ref{Ham_HIl_final}) the terms
that are responsible for the simultaneous creation/annihilation of
exciton-exciton and exciton-photon pairs. Then, using expressions
(\ref{HamIl_HamIv}), (\ref{Ham_Hv}) and
(\ref{Ham_HIl_final})--(\ref{polarization_operator}), after some
algebra  we derive the effective  two--particle  Hamiltonian
\begin{eqnarray}
\label{two_body_Ham}
\widehat{{\cal H}}_{\mathrm{eff}}&=&\widehat{{\cal H}}_\mathrm{0}+\widehat{{\cal H}}_\mathrm{ph}+\widehat{{\cal H}}_\mathrm{I0}+\Delta{\widehat{{\cal H}}}\,,  \\
\noalign{\hbox{where}} \label{two_body_HIL}
\widehat{{\cal H}}_\mathrm{I0}&=&\hbar\sum\limits_q(g_q\hat{\sigma}_{+}\hat{c}_q+g^*_q\hat{\sigma}_{-}\hat{c}^\dag_q) \\
\noalign{\hbox{and}}  \Delta \widehat{{\cal H}}&=&\frac{4\pi}{
V}\bm{\mu}(\tilde{\underline{N}}
\bm{\mu})(\hat{\sigma}_{-}\langle\hat{\sigma}_{+}
\rangle+\hat{\sigma}_{+}\langle\hat{\sigma}_{-} \rangle)
\,.\label{DeltaH}
\end{eqnarray}
where $g_{q}=-i\bm{ \mu}{\bf e}_{q}\sqrt{2\pi \omega _{k}/
\hbar\Omega }\exp(i{\bf kr}_{c})$ is the coupling factor for
photons and carriers in the QD and ${\bf r}_{c}$ is the
radius--vector of the QD geometrical center. The depolarization
tensor is given by
\begin{equation}
\label{Dep_Tens} \tilde{\underline{N}}=-\frac{V}{4\pi }
    \int\limits_{V}\!\!\int\limits_{V}|\xi(\mathbf{r})|^2\,|{\xi}(\mathbf{r}')|^2
    \underline{G}({\bf r}-\mathbf{r}')
   d^3\mathbf{r}\,d^3\mathbf{r}'\,.
 \end{equation}
Noted that the resulting Hamiltonian (\ref{two_body_Ham})
coincides with that obtained in Ref. \onlinecite{maxim_pra_02} in
independent way.

\subsection{Equations of motions } 
\label{sec:mot_eq}

Let $|\widetilde{\psi}(t)\rangle$ be a
wavefunction  of a QD interacting with quantum light. In the
interaction representation the system is described by the
Schr\"odinger equation
\begin{equation}
i \hbar\frac{\partial |\psi \rangle }{\partial t}=\widehat{{\cal
H}}_{\rm{int}}|\psi \rangle\,, \label{Schrod}
\end{equation}
with $|\psi(t)\rangle =\exp[i(\widehat{{\cal
H}}_{\rm{0}}+\widehat{{\cal H}}_{\rm{ph}})
t/\hbar]|\widetilde\psi(t)\rangle$ and $\widehat{{\cal
H}}_{\rm{int}}=\exp[i(\widehat{{\cal H}}_{\rm{0}}+\widehat{{\cal
H}}_{\rm{ph}})t/\hbar](\widehat{{\cal
H}}_\mathrm{I0}+\Delta\widehat{{\cal H}}) \exp[-(i\widehat{{\cal
H}}_0+\widehat{{\cal H}}_{\rm{ph}})t/\hbar]$. We represent the
wavefunction $|\psi(t)\rangle$ by the sum as follows
\begin{equation}
|\psi(t)\rangle=\sum_{\{n_{k}\}\geq{0}}\left[A_{\{n_{k}\}}(t)|e\rangle+B_{\{n_{k}\}}(t)|g\rangle\right]|\{n_{k}\}\rangle\,,
\label{wavefunct}
\end{equation}
where $A_{\{n_{k}\}}(t)$ and $B_{\{n_{k}\}}(t)$ are coefficients to
be found, $|\{n_{k}\}\rangle$ denotes the multimode field state with
$n$ photons in $k$ mode; $|\{0_k\rangle\}$ is the wavefunction of
the vacuum state of electromagnetic field; $|e \rangle$ and
$|g\rangle$ are the wavefunctions of the QD ground and excited
states, respectively. By inserting the relation (\ref{wavefunct})
into the Schr\"odinger equation (\ref{Schrod}), after some
manipulations we arrive at the system of equations of motion
\begin{eqnarray}
\label{eq_mot_bas0} i\frac{dA_{\{m_{l}\}} }{dt}&=&\Delta\omega
B_{\{m_{l}\}}\sum\limits_{\{n_{q} \}}A_{\{n_{q} \}}B_{\{n_{q}
\}}^*\cr
\rule{0in}{5ex}&+&\sum_{q}g_{q}\sqrt{m_{q}+1}B_{\{m_{l}+\delta_{lq}\}}e^{i(\omega_{0}-\omega_q)t}
\,,\\\label{bas0} i\frac{dB_{\{m_{l}\} }}{dt}&=&\Delta\omega
A_{\{m_{l}\}}\sum\limits_{\{n_{q} \}}A_{\{n_{q} \}}^*B_{\{n_{q}
\}}\cr
\rule{0in}{5ex}&+&\sum_{q}g^{*}_{q}\sqrt{m_{q}}A_{\{m_{l}-\delta_{lq}\}}e^{-i(\omega_{0}-\omega_q)t}
\,,~~\nonumber
\end{eqnarray}
with
\begin{equation}
\Delta\omega=\frac{4\pi}{\hbar{}V}{\bm \mu}(\widetilde{{
\underline N}}{\bm \mu}) \,, \label{dep_shift}
\end{equation}
as the local--field induced depolarization shift \cite{Slepyan_99a,maxim_pra_02}. Here $\omega_0=(\varepsilon_{\mathrm{e}}-\varepsilon_{\mathrm{g}})/\hbar$ is exciton transition frequency.  It can easily be shown that system (\ref{eq_mot_bas0}) satisfies the conservation
law
\begin{equation}
\label{conserv_law} \frac
{d}{dt}\sum_{\{n_{k}\}}(|{A}_{\{n_{k}\}}|^{2}+|{B}_{\{n_{k}\}}|^{2})=0\,.
\end{equation}

The system (\ref{eq_mot_bas0}) allows analyzing the interaction between QD and electromagnetic field of an arbitrary spatial
configuration and arbitrary polarization. Letting the coefficients $g_q(t)$
to be adiabatically slow--varying functions we can apply  (\ref{eq_mot_bas0}) to QDs exposed to electromagnetic pulse.

\subsection{Single--mode approximation}


Among different physical situations described by Eqs. (\ref{eq_mot_bas0}) the single-mode excitation is of special interest. Indeed, such a case corresponds, for instance, to the light
--QD interaction in a microcavity with a particular mode resonant with the QD exciton. Owing to the high Q-factor, the strong light--QD coupling regime is feasible in microcavity  providing numerous potential applications of such systems \cite{Scully,Michler_book}.

For the case of a spherical QD interacting with single--mode light,
only the components
$|\{n_{k}\}\rangle=|0_1,0_2,\dots{n_q}\dots{0_k}\dots\rangle=|n_q\rangle$
with $q$  as the number of interacting mode are accounted for in the wavefunction (\ref{wavefunct}).
Then, omitting for shortness the mode number, the system (\ref{bas0}) is reduced to
\begin{eqnarray}
\label{eq_mot_bas}
    i\frac{d{A}_{n}}{dt}&=&
      \Delta\omega {B}_{n}\sum\limits_{m}
      {A}_{m}{B}_{m}^*\cr \rule{0in}{4ex}&+&
      g\sqrt{n+1}{B}_{n+1}e^{i(\omega _{0}-\omega )t}\,,
      \label{bas1}\\ \rule{0in}{4ex}
    i \frac{d{B}_{n+1}}{dt} &=& \Delta\omega
      {A}_{n+1}
      \sum\limits_{m}{A}_{m}^*{B}_{m}\cr \rule{0in}{4ex}&+&
    g^*\sqrt{n+1}{A}_{n}
      e^{-i(\omega _{0}-\omega )t}\,. ~~\nonumber
\end{eqnarray}
Note that the conservation law
(\ref{conserv_law}) holds true for Eqs. (\ref{bas1}) with the substitution
$\{n_k\}\rightarrow{n}$.
Equations (\ref{eq_mot_bas0}) and (\ref{bas1}) govern  the time
evolution of the QD driven by quantum light.

\section{Free motion} 
\label{sec:free_mot}

The free motion regime implies neglect the QD--electromagnetic
field interaction and thus imposes the condition $g=0$ on Eqs.
(\ref{eq_mot_bas}). Wavefunction of the noninteracting QD and
electromagnetic field is factorized thus allowing analytical
solution of (\ref{eq_mot_bas}) in the form of
\begin{eqnarray}
\displaystyle A_{n}(t)=C_{n}A(t)\,,\qquad
\displaystyle B_{n}(t)=C_{n}B(t)\nonumber\,,
\end{eqnarray}
where $C_{n}$ are arbitrary constants satisfying the
normalization condition $\sum_n|C_n|^2=1$. In that case the system
(\ref{eq_mot_bas}) is reduced to the exactly integrable
form
\begin{eqnarray}
\displaystyle i\frac{dA}{d t} =
    \Delta \omega A|B|^{2}\,,  \qquad
    \displaystyle i\frac{dB}{d t}=
   \Delta\omega B|A|^{2}\,,
\label{eqs_fm}
\end{eqnarray}
and its solution is given by
\begin{eqnarray}
\displaystyle A(t)=a_0e^{-i\Delta\omega|b_0|^2 t}\,,  \quad
    \displaystyle B(t)=b_0e^{-i\Delta\omega|a_0|^2 t}\,.
\label{sol_eqs_fm}
\end{eqnarray}
Here $a_0$ and $b_0$  are arbitrary constants satisfying the
condition $|a_0|^2+|b_0|^2=1$. This solution describes a
\textit{correlated} motion of the electron-hole pair resulted from
the local field--induced self-polarization of the QD. Thus, a
quasi--particle with the wavefunction
\begin{eqnarray}
\label{wavefunct_fm} |\widetilde{\psi}(t)\rangle
=A(t){e^{-i\omega_et}}|e\rangle +B(t){e^{-i\omega_gt}}|g\rangle\,,
\end{eqnarray}
appears in the QD. It can easily be shown that the state
(\ref{wavefunct_fm}) satisfies the energy and probability
conservation laws. The inversion, which is defined as the
difference between the excited--state and the ground-state
populations of the QD exciton,  for the wavefunction
(\ref{wavefunct_fm}) remains constant in time:
$w=|A(t)|^2-|B(t)|^2\equiv |a_0|^2-|b_0|^2$,  whereas this state
is generally non-stationary. The quasi--particle lifetime, which
is not included in our model, can be estimated by
$\tau_\mathrm{sp}\sim 1/\Gamma_\mathrm{sp}$, where
$\Gamma_\mathrm{sp}$ is the QD--exciton spontaneous decay rate.
For realistic QDs $\Delta\omega\gg 1/\Gamma_\mathrm{sp}$
[\onlinecite{maxim_pra_02}]. Consequently, the state
$|\widetilde{\psi}(t)\rangle$ can be treated as stationary within
the range $1/\Delta\omega\ll{}t\ll 1/\Gamma_\mathrm{sp}$.

The macroscopic polarization of the QD is described by
\begin{eqnarray}
\langle{\widehat{\bm{\mathcal{P}}}}\rangle=\langle\psi|
\widehat{\bm{\mathcal{P}}}|\psi\rangle=\frac{1}{V}{\bm
\mu}a_0b^*_0 e^{-i(\omega_0-\delta')t}+\rm{c.c.}\,,
\label{polarization_fm}
\end{eqnarray}
where the parameter $\delta'={w}\Delta\omega$ plays the role of
the self--induced detuning, which depends on the state occupied by
the exciton and on the depolarization shift. Thus, as follows from
(\ref{polarization_fm}), the local field--induced  depolarization
shift ($\Delta\omega\neq 0$) dictates  the
\textit{non-isochronism} of the polarization oscillations, i.e.
the dependence of the oscillations frequency on its amplitude.
This mechanism also influences the Rabi oscillations in the
system: the smaller $\delta'$ the larger Rabi oscillations
amplitude; such a behavior  was observed experimentally in Ref.
\onlinecite{mitsumori_05}.

Since the inversion $w$ lies within the range $-1\leq{w}\leq{1}$,
the frequency $\omega_p$  of polarization oscillations in
(\ref{polarization_fm})  may vary in the limits
$\omega_0-\Delta\omega\leq\omega_p\leq\omega_0+\Delta\omega$. On
the contrary, when $\Delta\omega=0$  the polarization oscillates
with the fixed frequency $\omega_0$. At first glance, it seems
that the discrete level is transformed into $2\Delta\omega$ band.
However this is not the case. Indeed, the concept of the band
structure corresponds to linear systems where any arbitrary state
is a superposition of eigenmodes with different frequencies. As
different from that, the electron--hole correlation arises from
the nonlinear motion of the particles in a self-consistent field.
Consequently, in the presence of light--QD interaction ($g\neq
0$), the exciton motion can not be described by a simple
superposition of different partial solutions like
(\ref{sol_eqs_fm}), but has significantly more complicate
behavior. In particular,  in the strong coupling regime there exist two oscillatory
regimes with drastically different characteristics separated by
the bifurcation, see Sec. \ref{sec:rabioscillations}.

\section{The weak--field approximation}
\label{sec:weak_f_lim}

Consider a ground--state QD be exposed to
an arbitrary state of quantum light $\sum_{
m_l}\beta_{\{m_l\}}|\{m_l\}\rangle$, where $\beta_{\{m_l\}}$ are
arbitrary complex--valued coefficients satisfying the condition
$\sum_{\{m_l\}}|\beta_{\{m_l\}}|^2=1$. Then, the initial conditions for
Eqs. (\ref{eq_mot_bas0}) are given by
\begin{equation}
A_{\{m_l\}}(0)=0,\,\,\,  B_{\{m_l\}}(0)=\beta_{\{m_l\}}\,.
\label{init_cond_arb_field}
\end{equation}
In the linear regime with respect to the electromagnetic field
which realized when $g_q\rightarrow 0$ we can assume
$B_{\{m_l\}}(t)\approx{}\beta_{\{m_l\}}={\rm{const}}$, i.e. the
analysis is restricted to the time interval much less than the
relaxation time of the given exciton state. Then, the system
(\ref{bas0}) is reduced to
\begin{eqnarray}
\frac{dA_{\{m_{l}\}}
}{dt}&=&-i\Delta\omega\sum\limits_{\{n_q\}}A_{\{n_q\}}\beta_{\{m_{l}\}}\beta_{\{n_{q}\}}^*\cr
\rule{0in}{5ex}&-&\sum\limits_{q}g_{q}\sqrt{m_{q}+1}\beta_{\{m_{l}+\delta_{lq}\}}e^{i(\omega_{0}-\omega_q)t}
\,.
\label{eq_mot_weak_field}
\end{eqnarray}
For further analysis we  rewrite (\ref{eq_mot_weak_field}) in
more convenient matrix notation:
\begin{equation}
\frac{d{\bf a}(t)}{d t}=-i\Delta\omega{\bm \rho}{\bf a}(t)-{\bf
f}(t)\,, \label{eq_mot_weak_field_matrix_form}
\end{equation}
where ${\bf a}(t)$, ${\bf f}(t)=\sum_{q}{\bf
f}_q{e^{i(\omega_0-\omega_q)t}}$ and ${\bm\beta}$ are the columnar
matrices  (vectors) and ${\bf a}(t)= (A_{\{m_l\}} )$. Vectors ${\bf
f}_q$  and ${\bm\beta}$ are defined analogously through the elements
$f^{\{ m_{l} \}}_q=ig_q\sqrt{m_q+1}\beta_{\{m_l +\delta_{ql}  \}}$
and $\beta_{\{m_{l}\}}$, respectively; $\delta_{ql}$ is  the
Kroneker symbol. The quantity ${\bm
\rho}={\bm\beta}{\bm\beta}^{\dag}$ is the density matrix of quantum
light interacting with the QD. This matrix corresponds to the pure
state  of electromagnetic field and satisfies the conditions ${\bm
\rho}^n={\bm \rho}$ ($n\geq1$) and ${\bm \rho}^0=\mathbf{I}$ (where
$\mathbf{I}$ is the unit matrix). Integration of
(\ref{eq_mot_weak_field_matrix_form}) imposed by initial conditions
(\ref{init_cond_arb_field}) allows us to find the analytical
solution
\begin{equation}
{\bf a}(t)=-\int\limits_0^t{e^{-i{\bm \rho}\Delta\omega(t-t')}{\bf
f}(t')dt'}\,. \label{sol}
\end{equation}
Using the truncated Taylor expansion \cite{ref02}
\begin{eqnarray}
e^{-i{\bm \rho}\Delta\omega(t-t')} =\mathbf{I}-{\bm
\rho}(1-e^{-i\Delta\omega(t-t')}) \,,\nonumber
\end{eqnarray}
after some standard manipulations we obtain
\begin{eqnarray}
{\bf a}(t)&=&-i\sum\limits_{q}
    \left[({\bm \rho}-\mathbf{I}){\bf f}_q
\frac{1-e^{i(\omega_0-\omega_q)t} }{\omega_0-\omega_q-i0}\right.
\cr \rule{0in}{5ex}
    &-& \left.\bm{\rho}{{\bf
f}_q}e^{-i\Delta\omega{t}}\frac{(1-e^{i(\omega_0-\omega_q+\Delta\omega)t})}{\omega_0-\omega_q+\Delta\omega-i0}\right]\,.
\label{sol_full}
\end{eqnarray}
This expression describes the interaction between the QD and an
arbitrary state of quantum field in the weak--field limit.

Now let us calculate the QD effective scattering cross-section
defined by
\begin{equation}
\sigma(\infty)=\lim_{t\rightarrow{\infty}}\frac{d}{dt}|{\bf
a}(t)|^2 \,. \label{cross_section}
\end{equation}
Substituting (\ref{sol_full}) into
(\ref{cross_section}), after some algebra we arrive at
\begin{equation}
\sigma(\infty)=2\pi\sum_{\nu=1,2}\sum_{q}|{\bf
C}_{\nu{q}}|^2\delta{(\omega_{\nu}-\omega_{q})}\, ,
\label{cross_section_final}
\end{equation}
where  ${\bf C}_{1q}=(\bm{\rho}-\mathbf{I}){\bf f}_{q}$ and ${\bf
C}_{2q}=\bm{\rho}\,{\bf f}_{q}$, $\omega_{1}=\omega_{0}$. The
equation obtained demonstrates a fine structure of the absorption
in a QD exposed to quantum light with arbitrary statistics.
Instead of a single line with the exciton transition frequency
$\omega_{1}=\omega_{0}$, a doublet appears with one component
shifted to the blue: $\omega_{2}=\omega_{0}+\Delta\omega$. The
intensities of components are completely determined by the quantum
light statistics.  The same peculiarity has been revealed in Ref.
\onlinecite{maxim_pra_02} in a particular case of a QD exposed to
the single--mode Fock qubit. Obviously, this result directly
follows from (\ref{cross_section_final}) if we retain in this
equation only terms corresponding the mode considered.  The
single--mode Fock qubit is a superposition of two arbitrary Fock
states with fixed number of photons and is described by the
wavefunction (the mode index $q$ is  omitted)
\begin{eqnarray}
|\psi\rangle=\beta_{N}|N\rangle +\beta_{N+1}|N+1\rangle\,.
\label{Fock_qubit}
\end{eqnarray}
Consequently, at $N\geq 1$ the nonzero components of the vector
${\bf f}_{q}$  are $f^{(N-1)}=ig\sqrt{N}\beta_{N}$ and
$f^{(N)}=ig\sqrt{N+1}\beta_{N+1}$ . It can easily be found that in
that case Eq. (\ref{cross_section_final}) is reduced to Eq. (61)
from Ref. \onlinecite{maxim_pra_02}. In the case  $N=0$ the only
nonzero component $f^{(0)}=ig\beta_{1}$ survives in
(\ref{cross_section_final}) thus reducing this equation to Eq. (67)
from Ref. \onlinecite{maxim_pra_02}. For the single Fock state
$|\psi\rangle=|N\rangle$ and we obtain ${\bf C}_2=0$, i.e., in
agreement with Ref. \onlinecite{maxim_pra_02}, only shifted spectral
line is presented in the effective cross--section.

For the QD exposed to coherent states the recurrent formula
$\beta_{n+1}=\sqrt{\langle{n}\rangle/(n+1)}\beta_n$ can easily be
obtained, where $\langle{n}\rangle$ stands  for the photon number
mean value. Using this formula the relation $\bm{\rho} {\bf f}= {\bf
f}$ can be obtained, which gives ${\bf C}_1=0$. Thus, for this case
only shifted line in the effective cross section is manifested.
Analogous result has been reported in \onlinecite{maxim_pra_02} for
a QD driven by classical electromagnetic field. Such a coincidence
can be treated as the manifestation of well known concept: among a
variety of quantum states of light the coherent states are most
close to classical electromagnetic field.

\section{The quantum light--QD strong coupling regime. Rabi oscillations}
\label{sec:rabioscillations}  In the case when the coupling constant
is comparable in value with the exciton decay rate, the
linearization with respect to electromagnetic field utilized in the
previous section is no longer admissible. The strong coupling regime
can be realized by combining the QD with a high-Q microcavity and
one of its manifestation is the  Rabi oscillations of the levels
population in  a two--level system exposed to a strong
electromagnetic wave. In this section we utilize the system of
equations (\ref{eq_mot_bas}) for investigation of the interaction
between QD and single--mode quantum light. The incident field
statistics is exemplified by coherent states and Fock qubits.  In
the standard JC model the Rabi oscillations picture is characterized
by the following parameters\cite{Scully}: (i) the coupling constant
$g$; (ii) the frequency detuning $\delta=\omega_0-\omega$; (iii) the
initial distribution of photonic states in quantum light and (iv)
the average photon number $\langle{n}\rangle$. Accounting  for the
local field correction supplements this set with the new parameter,
the depolarization shift $\Delta\omega$. For convenience, we
introduce the depolarization shift by means of the
parameter\cite{magyar04} $\xi=\Omega_{\langle n
\rangle}/\Delta\omega$ which compares the shift with the Rabi
frequency $\Omega_{\langle n
\rangle}=2|g|\sqrt{\langle{n(0)}\rangle}$. To understand the
dynamical property of the Rabi effect we have investigated the
following four physical characteristics: the QD--exciton inversion
\begin{equation}
\label{inversion_def}
w(t)=\sum\limits_{n=0}(|A_{n}(t)|^2-|B_{n}(t)|^2)\,,
\end{equation}
the time evolution of the distribution of photonic states
\begin{equation}
\label{population_distribution}
p(n,t)=|A_{n}(t)|^2+|B_{n}(t)|^2\,,
\end{equation}
and the normally ordered normalized time--zero second order
correlation function of the driving field:

\begin{eqnarray}
g^{(2)}(t)&=&
    \frac{\langle\widehat{c}^\dag(t)\widehat{c}^\dag(t)\widehat{c}^{}(t)\widehat{c}^{}(t)\rangle}
    {{\langle\widehat{c}^\dag(t)\widehat{c}^{}(t)\rangle}^2}
    \cr\rule{0in}{5ex}
    &=&\sum\limits_{n=0} n(n-1)p(n,t)/{[\sum\limits_{n=1} n p(n,t)]^2}\,.
\label{corr_funct_Shrod}
\end{eqnarray}
The temporal evolution of the inversion can be detected in
pump--probe experiments \cite{kamada,htoon},  while the photonic
state distribution (\ref{population_distribution}) is measurable in
quantum nondemolition experiments with atoms \cite{Brune_pra92}.
\begin{figure}[th]
\begin{center}
\includegraphics[width=3.3in]{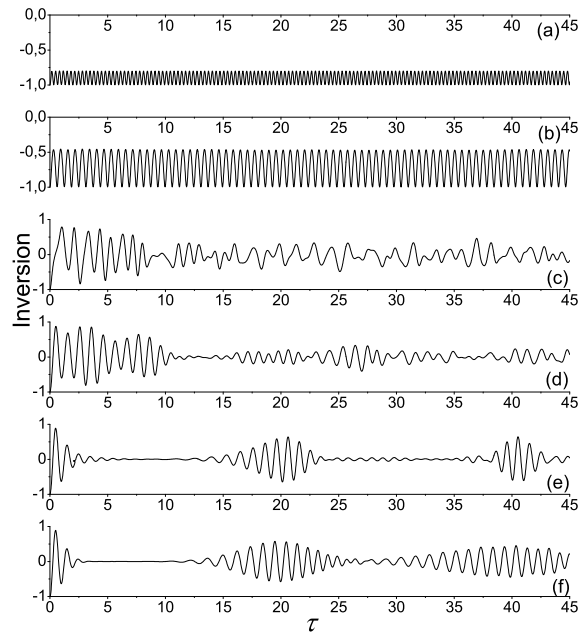}
\end{center}
\caption{Rabi oscillations of the inversion for a QD exposed to
the coherent state of light with $\langle{n(0)}{\rangle=9}$ and
$\xi=0.2$ (a),  $\xi=0.49$ (b), $\xi=0.53$ (c),  $\xi=1.2$ (d),
 $\xi=3.5 $ (e), $\xi=18.0$ (f).  } \label{fig_cs1_inv}
\end{figure}

\subsection{Coherent states excitation}
\label{sect:cohst}

\begin{figure}[th]
\begin{center}
\includegraphics[width=3.0in]{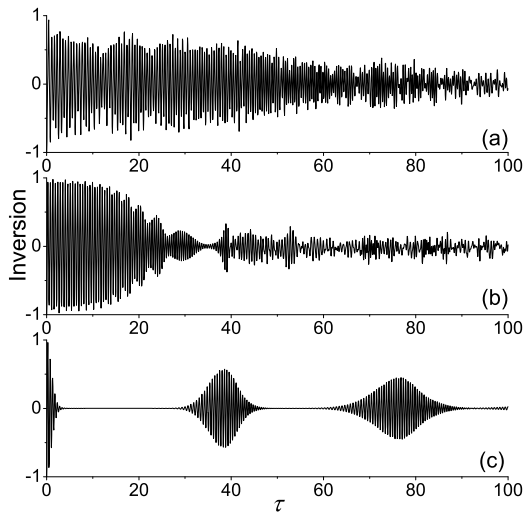}
\end{center}
\caption{Rabi oscillations of the inversion for a QD exposed to
the coherent state of light with $\langle{}n(0)\rangle=36$ and
$\xi=0.53$ (a), $\xi=1.2$ (b),  $\xi=18.0$ (c).}
\label{fig_cs2_inv}
\end{figure}
Let a ground--state QD be exposed to the elementary coherent
state of light $\sum_{n=0}^{\infty}F(n)|n\rangle$, where\cite{Scully}
$F(n)=\exp{[ -\langle{n(0)}\rangle
/{2}]}\langle{n(0)}\rangle^{{n}/{2}}/{\sqrt{n!}}$.
Then, initial conditions for Eqs. (\ref{bas1}) are given by
\begin{eqnarray}
B_{n}(0)=F(n) \,,\qquad A_{n}(0)=0 \,. \label{coh_st_init_cond}
\end{eqnarray}
Figures \ref{fig_cs1_inv}  and \ref{fig_cs2_inv} show calculations
of the inversion in a lossless QD as a function of the
dimensionless time $\tau =|g|t$  at the exact synchronism
($\delta=0$) for two different initial photon mean numbers
$\langle{n(0)}\rangle$ and  for several values of parameter $\xi$.
Our calculations demonstrate the appearance of two completely
different oscillatory regimes in the Rabi oscillations. The first
one manifests itself at  $\xi<0.5$ and is characterized by
periodic oscillations of the inversion within the range
$-1\leq{w(t)}<{0}$ (see Figs. \ref{fig_cs1_inv}a,
\ref{fig_cs1_inv}b). Thus, in this regime the inverted population
is unreachable. On the contrary, in the second regime, at
$\xi>0.5$, the inversion oscillates in the range
$-1\leq{w(t)}\leq{1}$ (Figs.
\ref{fig_cs1_inv}c--\ref{fig_cs1_inv}f and Fig.
\ref{fig_cs2_inv}). These two regimes  of the Rabi effect are
separated by the bifurcation which occurs at $\xi={0.5}$ for both
types of incident coherent states (compare Figs.
\ref{fig_cs1_inv}b and \ref{fig_cs1_inv}c).

In the limit $\xi\rightarrow{\infty}$ ($\Delta\omega\rightarrow
0$) the contribution of terms $O(\Delta\omega)$  in Eqs.
(\ref{bas1}) is small. The neglect these terms corresponds to
the elimination of the local field effect. In this case the system
(\ref{bas1}) is reduced to that follows from the standard JC model
\cite{Scully}, allowing thus the analytical solution:
\begin{eqnarray}
w(t)=&-&\sum\limits_{n=0}^{\infty}(B_{n+1}(0))^2\left[\frac{\delta^2}{\Omega_n^2}+\frac{4|g|^2(n+1)}{\Omega_n^2}\cos(\Omega_n
t)\right]
 \cr \rule{0in}{5ex}
    &-&|B_{0}(0)|^2\,, \label{JC_sol}
\end{eqnarray}
where $\Omega_n=\sqrt{\delta^2+4|g|^2(n+1)}$. The fundamental
effect predicted by this solution is the collapse--revivals
phenomenon in the time evolution of the inversion \cite{Scully}.
We have found that at $\xi\geq{18}$ the numerical simulation by
Eqs. (\ref{bas1}) leads to the same result as analytics
(\ref{JC_sol}), see Fig. \ref{fig_cs1_inv}f.
In  the case $\xi\rightarrow{0}$ the amplitude of Rabi oscillations
tends to zero and $w(t)\approx -1$.
\begin{figure}[ht]
\begin{center}
\includegraphics[width=3.0in]{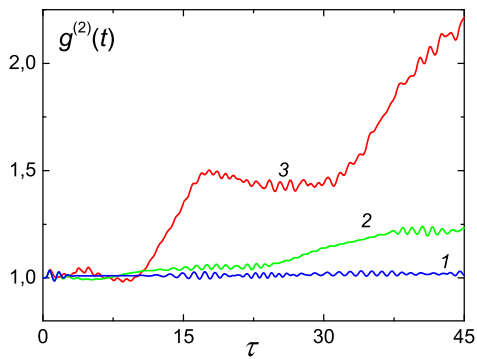}
\end{center}
\caption{The second order time--zero correlation function
$g^{(2)}(t)$ of QD exposed to the coherent state of light with
$\langle{n(0)}{\rangle{=}9}$ and $\xi=18.0$ (\textsl{1}),
$\xi=3.5$ (\textsl{2}), $\xi=1.2$ (\textsl{3}) }
\label{fig_cs1_cf}
\end{figure}

For a single QD imposed to classical light the appearance of two
oscillatory regimes in Rabi oscillations separated by the
bifurcation at $\xi=0.5$  has been predicted in Ref.
\onlinecite{magyar04}. Accordingly to [\onlinecite{magyar04}], the
region $\xi>0.5$  corresponds to periodic anharmonic oscillations of
the inversion. As different from that, in a QD exposed to quantum
light the collapse--revivals phenomenon takes place in this region,
see Figs. \ref{fig_cs1_inv}c--f and \ref{fig_cs2_inv}. As Figs.
\ref{fig_cs1_inv}f and \ref{fig_cs2_inv}c demonstrate, the collapse
and the revivals in the vicinity of the bifurcation are deformed and
turn out to be drastically different from those predicted by the
solution (\ref{JC_sol}). 
\begin{figure}[th]
\begin{center}
\includegraphics[width=3.0in]{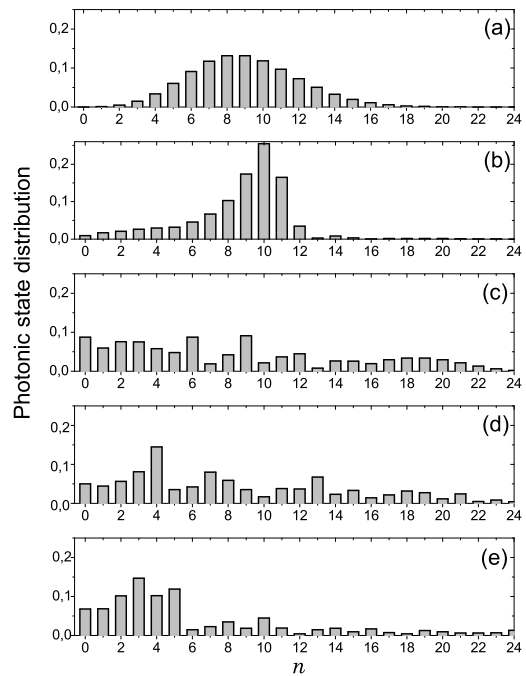}
\end{center}
\caption{Photonic state distribution in a QD exposed to coherent
light with $\langle n(0)\rangle = 9$ for $\xi=1.2$ and  $\tau=0$
(a), $\tau=8.5$ (b),  $\tau=14.5$ (c), $\tau=31$ (d),  $\tau=40$
(e).} \label{fig_cs1_pd}
\end{figure}
The collapse--revivals effect in the time evolution of the
inversion disappears completely in the range
$\xi<0.5 $ (see Figs. \ref{fig_cs1_inv}a--b),
where the Rabi effect picture turns out to be
identical to the case of QD excited by classical light \cite{magyar04}.

Let us estimate material parameters which provide observability of
the effects predicted.  For a spherical InGaAs QD with 6 nm radius
the dipole moment can be estimated\cite{ref05} as $\mu\approx 12$
Debye.  For this QD we obtain $\hbar\Delta\omega\approx 0.1$ meV.
Then, for the range of $\xi$ presented in Fig. \ref{fig_cs1_inv},
from $\xi=0.2$ to  $\xi=18$, we obtain $\hbar{\Omega_{\langle n
\rangle}}\approx 0.02$ and $\hbar{\Omega_{\langle n \rangle}}\approx
2$ meV, respectively. These values are of the same order as the
excitonic Rabi splitting measured in recent single QD spectroscopy
experiments, see Refs. \onlinecite{kamada,Yoshie_nature}. On the
other hand, Refs.
\onlinecite{birkedal_SM02,bayer_prb02,borri_prb05,silverman_apl06}
report the QD exciton linewidth $\Gamma_{{\rm hom}}$ of the order of
1 $\mu$eV below the temperature 10 K and  laying in the range 4 to
10  $\mu$eV  at $T=20$ K.  Thus, the precondition to observe the
strong coupling regime, $\hbar{\Omega_{\langle n \rangle}}\gg
\Gamma_{{\rm hom}} $, is  fulfilled for the given range of $\xi$.

An important feature of the Rabi effect in quantum light is the
variation of the light statistics during the interaction with a
quantum oscillator (QD). We shall characterize the variation  by the
second--order time--zero correlation function and the photonic state
distribution defined by Eqs. (\ref{corr_funct_Shrod}) and
(\ref{population_distribution}), respectively. These characteristics
at different $\xi$ are depicted in Figs. \ref{fig_cs1_cf} and
\ref{fig_cs1_pd}. One can see that at large $\xi\geq{}18$ the
function $g^{(2)}(t)$ oscillates around unit in agreement with the
standard JC model\cite{Scully}, see curve \emph{1} in Fig.
\ref{fig_cs1_cf}. The situation  is changed in the vicinity of the
bifurcation as it is presented by curves \emph{2} and \emph{3} in
that figure. As one can see, at $\xi=1.2$ the light statistics
becomes super--Poisonnian.   The correlation function $g^{(2)}(t)$
demonstrates the increase in time imposed to small-amplitude
oscillations. These oscillations correspond to regions of revivals
in the time evolution of the inversion (compare curve \emph{3} in
Fig. \ref{fig_cs1_cf} and Fig. \ref{fig_cs1_inv}d). It should be
noted that below the bifurcation threshold, at $\xi\leq{0.5}$,
$g^{(2)}(t)$ oscillates in the vicinity of unity too.

Fig. \ref{fig_cs1_pd} presents the photonic state distribution
$p(n,t)$ for different time points at the given $\xi=1.2$. The
figure illustrates consecutive transformation of the initially
Poissonian distribution (Fig. \ref{fig_cs1_pd}a) into the
super--Poisonnian one in the course of time, see Figs.
\ref{fig_cs1_pd}c--d. The transformation corresponds to the increase
in $g^{(2)}(t)$ illustrated by curve \emph{3} in Fig.
\ref{fig_cs1_cf}. Let us stress that the standard JC model predicts
the photon statistics remaining Poisson  as it takes place in our
case only  at large $\xi$ (curve \emph{1} in Fig. \ref{fig_cs1_cf}).
Our calculation also show that in the region $\xi<0.5$ the photon
distribution $p(n,t)$ does not depend on time and remains
Poissonian. The invariability of the coherent light statistics in
the limit $\xi\rightarrow{0}$ corresponds to the absence of the
component $\nu=1$ in Eq. (\ref{cross_section_final}).
\begin{figure}[th]
\begin{center}
\includegraphics[width=3.0in]{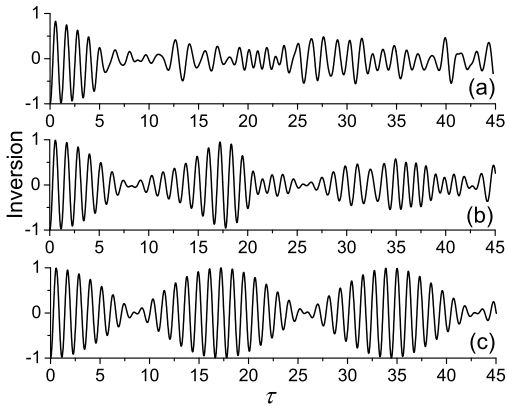}
\end{center}
\caption{Rabi oscillations of the inversion for the QD exposed to
the Fock qubit for $\xi=1.2$ (a), $\xi=8.0$ (b), $\xi=55.0$ (c).}
\label{fig_fq_inv}
\end{figure}

\subsection{QD interaction with Fock qubits} 
\label{sec:fock_qubit}

Let a ground-state QD interacts with electromagnetic field given by the Fock qubit
(\ref{Fock_qubit}). Then the initial
conditions for Eqs. (\ref{bas1}) are given by
\begin{eqnarray}
{A}_{n}(0)=0\,, \quad {B}_{n}(0)=\delta _{n,N}\beta_{N}+
    \delta _{n,N+1}\beta_{N+1}\,,
    \label{init_cond_Fock}
 \end{eqnarray}
Further we restrict ourselves to the Fock qubit with $N=6$ assuming
$\beta_{N}=\beta_{N+1}=\sqrt{1/2}$.

\begin{figure}[ht]
\begin{center}
\includegraphics[width=3.0in]{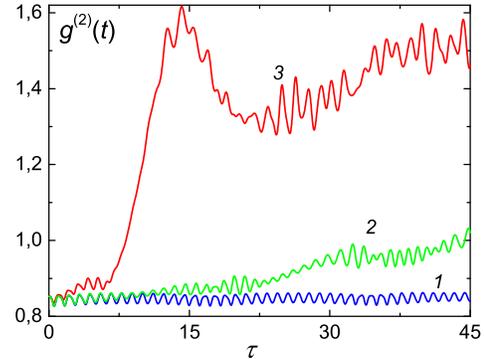}
\end{center}
\caption{Second order time-zero correlation function $g^{(2)}(t)$
of the QD exposed to the Fock qubit with $N=6$  and  $\xi=55.0$
$(\textsl{1})$, $\xi=8.0$ $(\textsl{2})$, $\xi=1.2$
$(\textsl{3})$.} \label{fig_fq_cf}
\end{figure}
\begin{figure}[ht]
\begin{center}
\includegraphics[width=3.2in]{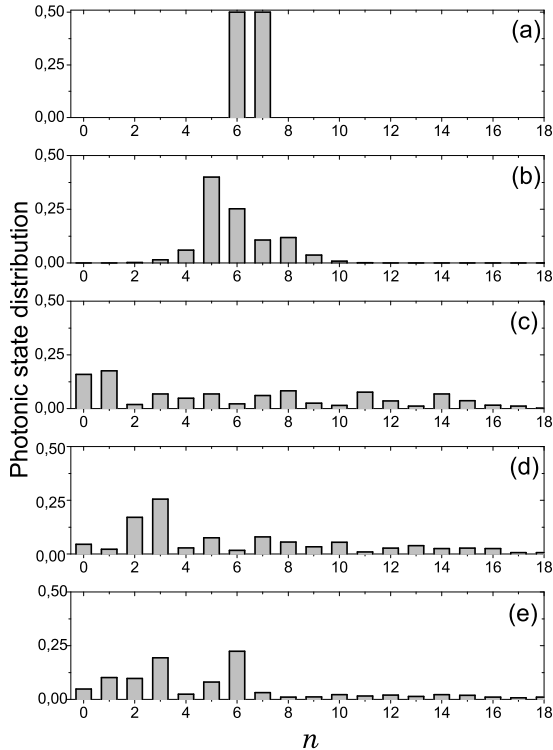}
\end{center}
\caption{Photonic state distribution of the QD exposed to the Fock
qubit for $\xi=1.2$ and  $\tau=0$ (a),   $\tau=5.19$ (b),
$\tau=14.5$ (c), $\tau=29.6$ (d),  $\tau=40.1$ (e).}
\label{fig_fq_pd}
\end{figure}
Calculations of time
evolution of the inversion for
different values of $\xi$ are shown in Fig. \ref{fig_fq_inv}. At large  $\xi$ (Fig.
\ref{fig_fq_inv}c) the local field effect is eliminated: In agreement with the
solution (\ref{JC_sol}) the
inversion exhibits harmonic modulation of the oscillation amplitude within the range $w\in[-1,1]$.
The oscillation frequency is equal to $(\Omega_N+\Omega_{N+1})/2$ while the frequency of the
modulation is given by  $(\Omega_N-\Omega_{N+1})/2$.
With the parameter $\xi$ decrease the modulation becomes non--harmonic.

The next two figures, \ref{fig_fq_cf} and \ref{fig_fq_pd},
illustrate the change of the light statistics due to
interaction with QDs. The time evolution of the correlation
function $g^{(2)}(t)$ is depicted in Fiq. \ref{fig_fq_cf} for
different values of $\xi$. Curve $\textsl{1}$ demonstrates that,
same as in Fig. \ref{fig_cs1_cf} and in agreement with the
standard JC model, at large $\xi$ ($\xi>55$) the correlation
function oscillates in the vicinity of its initial value.  Curve
$\textsl{3}$ shows transformation of the sub-Poissonian statistics
($g^{(2)}(t)<1$) into the super-Poissonian one ($g^{(2)}(t)>1$)
with a pronounced maximum followed by irregular oscillations.
Analogous but essentially smoothed  behavior is observed at larger
$\xi$ (curve  $\textsl{2}$).

Figure \ref{fig_fq_pd} presents calculations of the photonic state
distribution $p(n,t)$ for $\xi=1.2$ (curve $\emph{3}$ in Fig.
\ref{fig_fq_cf}) in different points of time. The standard JC model
when describes the interaction  of Fock qubit with two--level
system, predicts  variation  of those probabilities $p(n,t)$
(\ref{population_distribution}) that correspond to Fock states with
$n=N,N\pm 1$ numbers of photons. As different from that, the
incorporation of the local--field effect leads to the appearance in
the distribution of states $|n\rangle$ with photon numbers both
smaller and larger than present in the initial Fock qubit
(Fig.\ref{fig_fq_pd}a). Probabilities of these states are
redistributed (Figs. \ref{fig_fq_pd}b,c) with time, and light
statistics became irregular which signifies the transformation  of
the photonic state distribution (Figs.
\ref{fig_fq_pd}d,e). In turn this affects the Rabi oscillations
picture and the second order correlation function $g^{(2)}(t)$
(compare Figs. \ref{fig_fq_inv}a and \ref{fig_fq_cf}, curve
\textsl{3}).

\subsection{Vacuum Rabi oscillations}
\label{subsec:VacRO}

The vacuum Rabi oscillations characterize interaction of an
excited--state QD with electromagnetic vacuum. The initial
conditions for Eqs. (\ref{bas1}) in that case are given by
$B_n(0)=0, A_n(0)=\delta_{n,0}$. Numerical solution of this system
leads to the time--harmonic oscillations of the inversion. This
agrees with the analytical solution of the system at
$\Delta\omega=0$ given by the standard sinusoidal law\cite{Scully}:
\begin{equation}
w(t)=\frac{\delta^2}{\Omega_0^2}+\frac{4|g|^2}{\Omega_0^2}\cos(\Omega_0t)\,.
 \label{inv_vRO}
\end{equation}
The result can easily be understood from that the vacuum states, like
a single Fock state, have zero observable
electric field. Therefore, such states do not induce in QDs
observable polarization and, consequently, frequency shift. For
the same reason, zero frequency shift is inherent to
single--photon states, as it has been revealed under some
simplifying assumptions in Ref. \onlinecite{maxim_pra_02}.

\section{Discussion}
\label{sec:discussion}

The standard  JC model describing the interaction between
single--mode quantum electromagnetic field and two--level system
predicts  the collapse--revivals picture of the time evolution of
the level population. The  basic physical result of the analysis
presented in our paper is a significant modification of the Rabi
effect due to the local--field induced depolarization of a QD
imposed to quantum light. Two oscillatory regimes with drastically
different characteristics arise. In the first regime the time
modulation of the population (the collapse and revivals) is
suppressed and the QD population inversion is found negative. This
indicates that trajectories of charge carriers confined by the QD
occupy a finite volume in  the phase space. In the second
oscillatory regime the revivals appear, however they are found
deformed and significantly different from that predicted by the
standard JC model. The trajectories of charge carriers occupy the
entire phase space. Both regimes of oscillations demonstrate the
non--isochronous dependence on the coherent field strength.

Two regimes of the Rabi oscillations indicate the appearance of two
types of motion of the QD exciton. The first one is a superposition
of time-harmonic oscillations  with the Rabi frequencies
$\Omega_n=g\sqrt{N+1}$, while the second type is presented by the
frequency band $|\omega_p-\omega_0|<2\Delta\omega$ . The resulting
exciton motion is thus determined by a nonlinear superposition of
these two types of motion.

The first mechanism of the motion is conventional for the Rabi
effect; physically it originates from dressing  of the QD exciton by
the incident field photons. This type of motion dominates at large
strengths of the incident field, when
$\Omega_{\langle{n}\rangle}\gg\Delta\omega$. With the light--QD
coupling constant (and correspondingly $\Omega_{\langle{n}\rangle}$)
decrease, the role of the second type of motion grows in importance.
The reason for this is the electron--hole correlations resulting
from the exchange interaction. It can be interpreted as the QD
exciton dressing by virtual photons. This regime becomes dominating
in the comparatively weak fields, when
$\Omega_{\langle{n}\rangle}\leq\Delta\omega$. Thus, the reduction of
the threshold of the acting field strength needed for the Rabi
oscillations appearance, recently observed
experimentally\cite{mitsumori_05}, can be attributed  as a
local--field effect. The experiment\cite{mitsumori_05} has also
elucidated the non-isochronism of excitonic Rabi  oscillations that
can be treated as the  local--field effect as well; see the relation
(\ref{polarization_fm}) and numerical calculations reported in Ref.
[\onlinecite{magyar04}].

It should be noted that two oscillatory regimes in the Rabi effect
may appear in other quantum systems where additional interaction
mechanisms exist. As an example, consider a two--component
Bose--Einstein condensate with radio-frequency coupling of two
separate hyperfine states\cite{deconink_04}. Temporal evolution of
this system is governed by the coupled Gross--Pitaevskii equations,
which are similar to those derived in Sec. \ref{sec:mot_eq}. The
equations combine both linear and nonlinear couplings.  The linear
coupling constant characterizes the interaction between the system
and the electric field, while the nonlinear one accounts for the
interaction between intra- and inter- species of the
condensate\cite{deconink_04}. Dependently on ratio of the coupling
parameters, the Rabi oscillations between the condensate components
may  exhibit both well--ordered and chaotic  behavior
\cite{deconink_04}, similar to that depicted in Fig.
\ref{fig_cs1_inv}a,b. The formation of the
Bardeen--Cooper--Schrieffer state in the fermionic alkali gases
cooled bellow degeneracy [\onlinecite{barankov_04}] can serve as
another example. In that system, the time modulation of coupling
constant leads to the Rabi oscillations of the energy
gap\cite{barankov_ref} with two oscillatory regimes. The
trajectories of individual Cooper pairs occupy a finite volume in
the phase space in the first regime and the entire phase space in
the second one.

Now, let us turn to the weak--field case. The  the depolarization
induced local field is predicted to entail in a QD exposed to an
arbitrary photonic state a fine structure of the effective
scattering cross--section. Instead of a single line of the frequency
$\omega_0$, a duplet is appeared with one component shifted by a
value $\Delta\omega$. The shifted component is due to electron--hole
correlations, see Eq. (\ref{sol_eqs_fm}). The correlations change
the QD state and, consequently, provide the inelastic channel of the
light scattering. The elastic scattering channel is formed by light
states inducing zero observable polarization and, consequently, zero
frequency shift, such as Fock states, vacuum states, etc. Now we
take into account the relation $\langle{\bf \hat{P}}\rangle=
4\pi{\alpha }\langle{\bf \hat{E}}_0\rangle $, which couples
observable polarization and mean value of the incident field; the
quantity $\bf {\hat{E}}_0$ is defined through the relation
$\widehat{\bm{\mathcal{E}}}{}_0 (\mathbf{t})=
    \int_o^\infty
    \widehat{\mathbf{E}}_0(\mathbf{r},\omega)d\omega+ \rm{H.c}$. The
scalar coefficient $\alpha$ is the QD polarizability of a spherical
QD. Therefore, we conclude that the elastic scattering channel is
formed by incident field with zero mean value (incoherent component
of the electromagnetic field). Correspondingly, the coherent field
component is scattered through inelastic  channel. As follows from
the solution (\ref{cross_section_final}), the elastic channel is not
manifested for pure coherent light (the Glauber state).

Let us discuss now  the local--field induced alteration of the
quantum light statistics. As an example we consider the Fock qubit
(\ref{Fock_qubit}) as the incident field state. For the case
$\Delta\omega=0$ the photonic state distribution is given by
\begin{eqnarray}
\label{FQ_pn}
p(n,t)&=&\delta_{n,N-1}|A_{N-1}(t)|^2+\delta_{n,N}(|A_N(t)|^2\nonumber\\\rule{0in}{4ex}
&&+|B_N(t)|^2)
+\delta_{n,N+1}|B_{N+1}(t)|^2 \,,
\end{eqnarray}
where $A_n(t)$ and $B_n(t)$ are the exact solutions of Eqs.
(\ref{bas1}) at  $\Delta\omega=0$, see, e.g. Ref.
\onlinecite{Scully}. It is seen that the Fock states with photon
numbers  $n=N,N\pm 1$ are only presented in the distribution. The
probability amplitudes of these states  oscillate with the
corresponding Rabi frequencies, $\Omega_{n=N,N\pm 1}$.  In addition
to this set, extra Fock states with  both smaller and larger photon
numbers  are appeared in the photonic state distribution $p(n,t)$ as
the local--field effect, See Fig. \ref{fig_fq_pd}. Therefore, a
bigger number of Fock states than presented in the initial Fock
qubit, broaden the frequency spectrum of Rabi oscillations,
providing thus the chaotic time evolution of the inversion.

It should be noted  that the variation in the quantum light
statistics occurs even in the weak--field limit $g\rightarrow{0}$.
To illustrate that, consider the observable polarization of the QD
exciton defined in the time domain as
\begin{eqnarray} \langle{\widehat{\bm{\mathcal{P}}}}\rangle
&=&\frac{1}{V}{\bm
\mu^*}\sum\limits_{\{n_q\}}A_{\{n_q\}}B_{\{n_q\}}^*{\rm
e}^{-i\omega_0t}+\rm{c.c} \,. \label{discussions:Pol}
\end{eqnarray}
Using Eq. (\ref{sol_full}) we couple the polarization in the frequency domain with the complex--valued amplitude of the mean incident field:
\begin{eqnarray}
\langle{\bf \hat{P}}\rangle&=&\frac{|{\bm
\mu}|^2/\hbar{V}}{\omega_0-\omega+\Delta\omega-i0}\langle {\bf
\hat{E}}_0\rangle \,. \label{discussions:Pol2}
\end{eqnarray}
Then, after some simple manipulations we express
the quantum fluctuations of the QD polarization by
\begin{eqnarray}
{\bf \hat{P}}-\langle{\bf \hat{P}}\rangle&=&\frac{|{\bm
\mu}|^2/\hbar{V}}{\omega_0-\omega-i0}\left({\bf \hat{E}}_0-\langle
{\bf \hat{E}}_0\rangle\right)  \,. \label{discussions:Pol3}
\end{eqnarray}
From Eqs. (\ref{discussions:Pol2}) and (\ref{discussions:Pol3})
follows that the QD electromagnetic responses to  the mean electric
field and to its quantum fluctuation are different: the response
resonant frequency is shifted by the value $\Delta\omega$ in the
former case [relation (\ref{discussions:Pol2}) ] and remains
unshifted in the later one, as given by relation
(\ref{discussions:Pol3}). This  indicates  that the effective
polarizability  of a QD  is an operator in the space of quantum
states of light. It should be pointed out that this property is
responsible for the alteration of the photonic state distribution in
the weak--field regime and is entirely a local--field effect. Note
that the notions ''strong (weak) coupling regime'' and ''strong
(weak) field regime'' are not identical as applied to QDs. To
illustrate this statement,  we express from Eqs.
(\ref{discussions:Pol2}) and (\ref{discussions:Pol3}) the
polarization operator in the weak--field limit:
\begin{eqnarray}
{\bf \hat{P}}&=&\frac{|{\bm
\mu}|^2/\hbar{V}}{\omega_0-\omega-i0}\left({\bf \hat{E}}_0-
\frac{\Delta\omega{\langle {\bf
\hat{E}}_0\rangle}}{\omega_0-\omega+\Delta\omega-i0}\right) \,.\,\,
\label{discussions:Pol4}
\end{eqnarray}
This equation is linear in the incident field but includes a term
quadratic in the oscillator strength, $O(|\bm{\mu}|^4)$.
Nonlinearity of that type violates the weak coupling regime. The
splitting of the QD-exciton spectral line dictated by Eq.
(\ref{discussions:Pol3}) is a manifestation of the strong
light-matter coupling in  the weak incident--field regime. Thus,
the light--QD interaction is characterized by two coupling
parameters, the standard Rabi--frequency and a new one, the
depolarization shift $\Delta\omega$.

\section{Conclusions}
\label{sec:conclusion}

In the paper we have developed a
theory of the electromagnetic response of a single QD exposed to
quantum light, corrected to the local--field effects.
The theory  exploits the
two-level model of QD with both linear and nonlinear coupling
of excited and ground states. The nonlinear coupling is provided by the local field influence.
Based on the microscopic Hamiltonian accounting for the electron--hole exchange
interaction, an effective two--body Hamiltonian
has been derived and expressed in terms of the incident electric
field, with a separate term responsible for the local--field impact. The
quantum equations of motion  have been formulated and
solved with this Hamiltonian for different types of the QD excitation,
such as Fock qubit, coherent state, vacuum state and arbitrary
state of quantum light.

For a QD exposed to coherent light we predict two oscillatory
regimes in the Rabi oscillations separated by the bifurcation. In
the first regime,  the standard collapse--revivals phenomenon  do
not reveal itself and the QD inversion is found negative.
In the
second regime, the collapse--revivals picture is found to be
strongly distorted as compared with that  predicted by the standard
JC model. The model developed can easily be extended to systems of
other physical  nature  exposed to a strong electromagnetic
excitation. In particular, we expect manifestation of the
local--field effects in  Bose--Einstein condensates
\cite{deconink_04} and  fermionic alkai gases cooled below the
degeneracy \cite{barankov_04}.

We have also demonstrated that the local--field correction alters
the light statistics even in the weak--field limit. This is because
the local fields  give rise to the inelastic scattering channel for
the coherent light component. As a result, coherent and incoherent
light components  interact with QD on different frequencies
separated by the depolarization  shift $\Delta\omega$. In other
words, the local fields eliminate the frequency degeneracy between
these components of the incident light.

Note that our model does not account for the  dephasing. Accordingly
to recent experimental measurements \cite{Borri_prb02,borri_prb05}
and theoretical estimates \cite{forstner_03,dizhu_05}, the
electron--phonon interaction is the dominant mechanism of the
dephasing in QDs.  Thus, the further development of the theory
presented requires this dephasing mechanism incorporation. A next
step is generalization of our model to multi--level systems. Among
them, the systems with dark excitons interacting with weak probe
pulse in the self--consistent transparency regime
\cite{fleischhauer_05} are of special interest.

In the paper we have considered an isolated QD. The generalization of
the theory developed to the case of  QD ensembles
(excitonic composites), such as self--organized lattices of
ordered QD molecules \cite{mano_04} and 1D--ordered (In, Ga)As QD
arrays \cite{lippen_04}, is of special interest. One can expect
that dipole--dipole interactions between QDs will manifest itself
in a periodic transfer of excited state between QDs
resulting thus in the collective Rabi oscillations
--- Rabi waves.

\acknowledgments

The work was partially supported through the INTAS under project 05-1000008-7801 and the Belarus Republican Foundation for Fundamental Research under project  F05-127. The work of S. A. M. was partially
carried out during the stay at the Institute for Solid State
Physics, TU Berlin, and supported by the Deutsche
Forschungsgemeinschaft (DFG). Andrei Magyarov
acknowledges the support from the INTAS YS fellowship
under the project 04-83-3607.

\appendix

\section{Depolarization shift for the spherical QD}

For a spherical QD, the formulae (\ref{dep_shift}) is reduced to
\begin{equation}
\label{a1} \Delta\omega=\frac{4\pi}{3 \hbar V}|{\bm  \mu} |^2 {\rm
Tr}( \widetilde{\underline{N}})      \,.
\end{equation}
Using the expression (\ref{DeltaH}) we obtain
\begin{eqnarray}
\label{a2} {\rm Tr}(
\widetilde{\underline{N}})&=&\frac{V}{4\pi}\int\limits_{V}\!\!|\zeta(\textbf{r}')|^2\int\limits_{V}\!\!|\zeta(\textbf{r})|^2\nabla^2\left(\frac{1}{|\textbf{r}-\textbf{r}'|}\right)d^3\textbf{r}d^3\textbf{r}'\cr
\rule{0in}{5ex}&=&V\int\limits_{V}\!\!\int\limits_{V}\!\!|\zeta(\textbf{r}')|^2\zeta(\textbf{r})|^2\delta(\textbf{r}-\textbf{r}')d^3\textbf{r}d^3\textbf{r}'\cr
\rule{0in}{5ex}&=&V\int\limits_{V}|\zeta(\textbf{r})|^4
d^3\textbf{r} \,.
\end{eqnarray}
Consider the 1s state. The normalized  wavefunction in this case
is given by \cite{haug_b94}
\begin{equation}
\label{a3} \zeta(\textbf{r})= \frac{1}{R\sqrt{2\rho}}J_{1/2}\left(
\pi\frac{\rho}{R}\right) \,,
\end{equation}
where $J_{\nu}(\dots)$ is the Bessel function, $\rho$ is the
spherical coordinate and $R$ is the QD radius. Inserting
(\ref{a3}) into (\ref{a2}) and integrating the resulting
expression, we derive the correction to the depolarization shift
(\ref{a1})
\begin{equation}
\label{a4} {\rm Tr}(
\widetilde{\underline{N}})=\frac{4\pi}{3}\int\limits_0^{\pi}\frac{\sin^4x}{x^2}dx\approx
2.81\,.
\end{equation}


\begin{thebibliography}{99}
\bibitem{Scully} M.  O.   Scully  and  M. S. Zubairy,
\textit{Quantum Optics} (University Press, Cambridge, 2001).

\bibitem{bimberg_b99} D. Bimberg, M. Grundmann, and N. N. Ledentsov, {\it Quantum dot heterostructures}. (John Wiley \&
Sons, Chichester, 1999).

\bibitem{Michler_book} {\it Single quantum dots, Topics  of applied
physics}, edited by P. Michler (Springer-Verlag Berlin,
Heidelberg, 2000)

\bibitem{kasevich} M. Kasevish, Science \textbf{298}, 1363 (2002).

\bibitem{lounis_05} B. Lounis and M. Orrit, Rep. Prog. Phys. \textbf{68}, 1129 (2005).

\bibitem{bratke} S. Brattke, B. T. H. Varcoe, and H. Walther, Phys.
Rev. Lett. \textbf{86}, 3534 (2001)

\bibitem{Law_96} C. K. Law and J. H. Eberly, \prl {\bf 76}, 1055 (1996).

\bibitem{Kilin_99} S. Ya. Kilin, Usp. Fiz. Nauk \textbf{169}, 507 (1999) [Phys. Usp. \textbf{42}, 435 (1999)].

\bibitem{zadkov} I. V. Bagratin, B. A. Grishanin, and V. N. Zadkov, Usp. Fiz. Nauk \textbf{171}, 625 (2001) [Phys. Usp. \textbf{44}, 597  (2001)].

\bibitem{wheeler} Y. A. Wheeler  and  W. H. Zurek, {\it Qunatum theory of
measurment}, (Princenton Univ. Press, 1983).


\bibitem{Jaynes_Cummings} E. T. Jaynes and F. W. Cummings, Proc. IEEE \textbf{51}, 89 (1963).


\bibitem{yang_04} Y. Yang, J. Xu, G. Li  and, H.Chen, Phys. Rev. A \textbf{69}, 053406 (2004).

\bibitem{lewenstein_94} M. Lewenstein, L. You, J. Cooper and, K.
Burnett, Phys. Rev. A \textbf{50}, 2207 (1994).

\bibitem{Zhang_94} Weiping Zhang and D. F. Walls, \pra
\textbf{49}, 3799 (1994).

\bibitem{dizhu_05} Ka-Di Zhu, Zhuo-Jie Wu, Xiao-Zhong Yuan, and Hang Zheng, \prb  \textbf{71}, 235312 (2005).

\bibitem{forstner_03} J. F\"orstner, C. Weber, J. Danckwerts, and A. Knorr, Phys. Rev. Lett.
\textbf{91}, 127401 (2003); J. F\"orstner, C. Weber,  J. Danckwerts,
and A. Knorr, Phys. Status. Solidi B, \textbf{238}, 419 (2003).


\bibitem{fleischhauer_05} M.   Fleischhauer,  A.  Immamoglu, and J. P. Marangos, Rev. Mod. Phys. \textbf{77}, 633 (2005).


\bibitem{Dung_02} H. T. Dung, L. Knoll and D.-G. Welsch, \pra \textbf{66}, 063810 (2002).

\bibitem{Unold_05} Th. Unold, K. Mueller, C. Lienau, Th. Elsaesser, and A. D. Wieck, \prl \textbf{94}, 137404 (2005).


\bibitem{sticvater}T.  H.   Stievater, Xiaoqin Li, D. G. Steel, D. Gammon, D. S. Katzer, D. Park, C. Piermarocchi, and L. J. Sham, Phys. Rev. Lett. \textbf{87}, 133603 (2001).

\bibitem{kamada} H. Kamada, H. Gotoh, J. Temmyo, T. Takagahara, and H. Ando, Phys. Rev. Lett. \textbf{87}, 246401 (2001).

\bibitem{htoon} H. Htoon, T. Takagahara, D. Kulik, O. Baklenov, A. L. Holmes, Jr., and C. K. Shih
Phys. Rev. Lett. \textbf{88}, 087401 (2002).

\bibitem{Zrenner_nature} A. Zrenner, E. Beham, S. Stufler, F. Findels, M. Bichler, G.
Abstreiter, Nature \textbf{418}, 612 (2002).

\bibitem{sticvater_rep} X. Li, Y. Wu, D. Steel, D. Gammon, T.  H.  Stievater, D. S. Katzer, D. Park, C. Piermarocchi, and L. J.
Sham, Science \textbf{301}, 5634 (2003).

\bibitem{mitsumori_05} Y. Mitsumori, A. Hasegawa, M. Sasaki, H. Maruki, and F. Minami
Phys. Rev. B \textbf{71}, 233305 (2005).


\bibitem{Schmitt_87}  S. Schmitt-Rink, D. A. B. Miller, and D. S. Chemla,
Phys. Rev. B  {\bf 35}, 8113 (1987).

\bibitem{Hanewinkel_97} B. Hanewinkel, A. Knorr, P. Thomas, and S.
W. Koch, Phys. Rev. B {\bf 55}, 13715 (1997).

\bibitem{Slepyan_99a} G. Ya. Slepyan, S. A. Maksimenko, V. P. Kalosha, J. Herrmann, N. N.
Ledentsov, I. L. Krestnikov, Zh. I. Alferov, and D. Bimberg, Phys.
Rev. B {\bf 59}, 12275 (1999).

\bibitem{Maksim_00a} S. A. Maksimenko, G. Ya. Slepyan, V. P. Kalosha, S. V.
Maly, N. N. Ledentsov, J. Herrmann, A. Hoffmann, D. Bimberg, and
Zh. I. Alferov, J. Electron. Mater.  \textbf{29}, 494 (2000).

\bibitem{Ajiki_02} H. Ajiki, T. Tsuji, K. Kawano, and K. Cho,
Phys. Rev. B {\bf 66}, 245322 (2002).



\bibitem{Goupalov_03} S. V. Goupalov, Phys. Rev. B {\bf 68}, 125311 (2003).

\bibitem{Slepyan_NATO_03} G. Ya. Slepyan \textit{et al.},  in  \textit{Advances in Electromagnetics
of Complex Media and Metamaterials}, edited by S. Zouhdi
\textit{et al.} (Kluwer, Dordrecht, 2003), p. 385.

\bibitem{Maksimenko_ENN} S. A. Maksimenko and G. Ya. Slepyan,
in \textit{Encyclopedia of Nanoscience and Nanotechnology}, edited
by J. A. Schwarz \textit{et al.} (Marcel Dekker, New York, 2004),
p. 3097.

\bibitem{Maksimenko_HN04} S.A. Maksimenko and G.Ya. Slepyan, in
\textit{The Handbook of Nanotechnology: Nanometer Structure
Theory, Modeling, and Simulation,} edited by A. Lakhtakia (SPIE
Press, Belingham, 2004), p. 145.

\bibitem{maxim_pra_02} G. Ya. Slepyan, S. A. Maksimenko, A. Hoffmann and D. Bimberg, Phys. Rev. A \textbf{66}, 063804 (2002).


\bibitem{magyar04} G. Ya. Slepyan, A. Magyarov, S. A. Maksimenko, A. Hoffmann, and D. Bimberg, \prb \textbf{70}, 045320 (2004).

\bibitem{ref01} This approach is often
utilized for the description of  d-d interactions in atomic optics,
see e.g. Ref. \onlinecite{Zhang_94}.


\bibitem{beresteckij} V. B. Berestetskii, E. M. Lifshitz, and L. P. Pitaevskii, \textit{ Quantum
Electrodynamics}, (Pergamon, Oxford, 1982).


\bibitem{hartree_rigourus} Applicability of the Hartree--Fock--Bogoliubov approximation to our problem was discussed in
previous publications under the phenomenological derivation of
Hamiltonian (\ref{two_body_Ham}), see Refs.
\onlinecite{maxim_pra_02,Maksimenko_ENN}.

\bibitem{Cho_b03} K. Cho, \textit{Optical Response of Nanostructures: Microscopic Nonlocal Theory }(Springer-Verlag, Berlin, 2003).

\bibitem{Chow_b99} W. W. Chow and S. W. Koch, \textit{Semiconductor-Laser Fundamentals:
Physics of the Gain Materials}, (Springer-Verlag, Berlin, 1999).


\bibitem{haug_b94} H. Haug and S. W. Koch, \textit{Quantum theory of the
optical and electronic properties of semiconductors}, (World
Scientific, Singapure 1994).

\bibitem{ref02} This expression is obtained
by i) the exponent expansion into Taylor series, ii) utilizing the
relations ${\bm \rho}^n={\bm \rho}$ ($n\geq1$) and ${\bm
\rho}^0=\rm{I}$ and iii) subsequent summations of the Taylor
series.


\bibitem{Brune_pra92} M. Brune,  S. Haroche, J. M. Raimond, L. Davidovich, and N. Zagury, \pra \textbf{45}, 5193 (1992).




\bibitem{ref05}  The dipole moment in the expression for the depolarization shift
(\ref{a1}) for InGaAs spherical QD can be estimated using the
experimental data of  Ref. [K.L. Silverman, R.P. Mirin, S.T.
Cundiff, A.G. Normann,   Appl. Phys. Lett. \textbf{82}, 25, (2003)]
where the value from 25 to 33 Debye were measured for the QDs with
average lateral dimensions from 30 to 40 nm. Taking into account the
theretical results concerning the dependence of  the transition
dipole moment on the QD surface area  and  shape [ A.
Th\"{a}nharrdt, C. Ell, G. Khitrova, H.M. Gibbs \prb  \textbf{65},
035327, (2002)], we can roughly estimate the dipole moment $\mu
\approx 12$ Debye for spherical QD with radius 6 nm.


\bibitem{Yoshie_nature} T. Yoshie, A. Scherer, J. Hendrickson, G. Khitrova, H. M. Gibbs, G. Rupper, C. Ell, O. B. Shchekin, and D. G. Deppe,  Nature \textbf{432}, 11 (2004).

\bibitem{birkedal_SM02} D. Birkedal, K. Leosson, and J. M. Hvam, Superlattices and
Microstructures \textbf{31}, 97 (2002).

\bibitem{Borri_prb02} P. Borri, W. Langbein, S. Schneider, U.
Woggon, R. L. Sellin, D. Ouyang, and D. Bimberg, Phys. Rev. B
\textbf{66}, 081306(R) (2002).

\bibitem{borri_prb05} P. Borri, W. Langbein, U. Woggon, V.
Stavarache, D. Reuter, A.D. Wieck \prb, \textbf{71}, 115328 (2005).

\bibitem{silverman_apl06} J. J. Berry, M. J. Stevens, R. P. Mirin, and K. L.
Silverman, \apl \textbf{88}, 061114 (2006).

\bibitem{bayer_prb02}  M. Bayer and A. Forchel,  \prb \textbf{65}, 041308(R) (2002).













\bibitem{deconink_04} B. Deconinck, P. G. Kevrekidis, H. E. Nistazakis, and D. J. Frantzeskakis, Phys. Rev. A  \textbf{70}, 063605 (2004).

\bibitem{barankov_04} R. A. Barankov, L. S. Levitov, and B. Z.
Spivak, Phys. Rev. Lett. \textbf{93}, 160401 (2004).

\bibitem{barankov_ref}  Rabi oscillations of the energy
gap correspond to the periodic creation and annihilation of the
Cooper pairs.


\bibitem{mano_04} T. Mano, R. Notzel, G. J. Hamhius, T. H. Eijkemans, and J. H. Wolter, J. Appl. Phys. \textbf{95}, 109 (2004).

\bibitem{lippen_04} T. V. Lippen,  R. Notzel, G. J. Hamhius, and J. H.
Wolter, \apl \textbf{85}, 118 (2004).




\end{thebibliography}
\end{document}